\documentclass{nature}
\usepackage{amsmath}
\usepackage{amssymb}
\usepackage[skip=0pt]{caption}
\usepackage[edtable]{lineno}
\usepackage{graphicx}
\usepackage[usenames]{color}
\usepackage{relsize}
\usepackage{printlen}
\usepackage{soul,xcolor}

\definecolor{Nathanblue}{rgb}{0.,0.24,0.51}

\definecolor{Seba}{RGB}{00,0,0}
\usepackage[colorlinks=true,
            linkcolor=magenta,
            urlcolor=blue,
            citecolor=magenta]{hyperref}

\renewcommand{\figurename}{{\bf Figure}}

\title{State-recycling and time-resolved imaging in topological photonic lattices}

\author{Sebabrata Mukherjee$^{1,}$\footnote{mukherjeesebabrata@gmail.com, \, $^{\dagger}$ngoldman@ulb.ac.be, \, $^{\ddagger}$r.r.thomson@hw.ac.uk}, Harikumar~K.~Chandrasekharan$^{1}$, Patrik \"Ohberg$^{1}$, Nathan Goldman$^{2, \dagger}$, \& Robert R. Thomson$^{1, \ddagger}$}

\begin{document}

\setstcolor{red} 

\maketitle

\begin{affiliations}
\item Scottish Universities Physics Alliance (SUPA), Institute of Photonics and Quantum Sciences (IPaQS), School of Engineering $\&$ Physical Sciences, Heriot-Watt University, Edinburgh, EH14 4AS, United Kingdom
\item Center for Nonlinear Phenomena and Complex Systems, Universit\'e Libre de Bruxelles, CP 231, Campus Plaine, B-1050 Brussels, Belgium
\end{affiliations}

\clearpage

\begin{abstract}
Photonic lattices - arrays of optical waveguides - are powerful platforms for simulating a range of phenomena, including topological phases. While probing dynamics is possible in these systems, by reinterpreting the propagation direction as ``time," accessing long timescales constitutes a severe experimental challenge. Here, we overcome this limitation by placing the photonic lattice in a cavity, which allows the optical state to evolve through the lattice multiple times. The accompanying detection method, which exploits a multi-pixel single-photon  detector array, offers quasi-real time-resolved measurements after each round trip. We apply the state-recycling scheme to intriguing photonic lattices emulating Dirac fermions and Floquet topological phases. {\color{Seba}In this new platform, we also realise a synthetic pulsed electric field, which can be used to drive transport within photonic lattices.}
This work opens a new route towards the detection of long timescale effects in  engineered photonic lattices and the realization of hybrid analogue-digital simulators.
\end{abstract}

\newpage

{\bf Introduction}\\
In the last decade, topological photonics has emerged as a promising field for the realisation and detection of exotic states of matter with topological properties~\cite{Lu2014,Khanikaev2017}. Building lattices for light has in particular allowed for the engineering of topological phases that have remained inaccessible in solid-state devices, such as Floquet topological phases~\cite{Kitagawa_walk,Rechtsman2013, Mukherjee2017, Maczewsky2017}, and has offered novel methods by which the geometry and topology of Bloch bands can be directly extracted~\cite{Mittal2016,Wimmer2017,Cardano2017}. Among the various photonic devices developed so far, {\it photonic lattices}, consisting of periodic arrays of coupled optical waveguides, provide a particularly
rich toolbox for the simulation of intriguing toy-models of topological phenomena~\cite{Rechtsman2013, Mukherjee2017, Maczewsky2017}.
Often realized utilising ultrafast-laser-fabrication techniques~\cite{Davis1996}, these engineered lattices allow for independent and dynamical control over the effective inter-site tunnelling and on-site potentials, and can be arranged into various geometries. Beyond topological effects, photonic lattices have also been exploited to investigate many other effects~\cite{christodoulides2003discretizing, Garanovich2012} including quantum correlations~\cite{Bromberg2009, Peruzzo2010} and the photonic Zeno effect~\cite{Biagioni2008}.

In the scalar-paraxial approximation, light propagation across a photonic lattice is governed by a Schr\"odinger-like equation~\cite{Garanovich2012}, where the propagation distance ($z$) plays the role of ``time" ($z\!\leftrightarrow\!t$). {\color{Seba}In current photonic lattice simulators, unlike fibre networks~\cite{regensburger2012parity}, the effective} time-evolution of a specific input state is measured over relatively short timescales, which are set by the maximum propagation distance $L\!\sim\!10$~cm of the fabricated lattices. This approach complicates, or even prevents, the observation of physical phenomena that are associated with slow dynamics, such as those emanating from weak effective inter-particle interactions~\cite{Creffield2004, Lumer_interactions} or weakly dispersive bands~\cite{Khomeriki2016}. In addition, it prevents the study of topological edge modes over long durations, and in particular topological interference effects~\cite{Heiblum_edge}.

Here, we propose to overcome this limitation by placing the photonic lattice in an optical cavity and recycling the optical state through the lattice multiple times.
After each cavity round-trip, the ``time-evolved" output state is then observed  using {pulsed excitation light and an}
advanced single-photon avalanche detector (SPAD) array~\cite{Richardson2009}, which facilitates independent  time-correlated single-photon counting for each mode of the photonic lattice. We demonstrate the operation of two types of cavities, and apply these to study the quasi-real-time evolution of pseudo-relativistic modes and Floquet anomalous topological edge modes over long distances. We also show how synthetic electric fields can be naturally introduced in this platform, hence offering a simple method by which transport can be driven within photonic lattices. In principle,  the output state could be finely modified after each round-trip, offering the possibility of engineering quantum walks, local dissipation, gauge fields and effective interaction effects in a (quasi-real-time) stroboscopic manner.

\begin{figure*}[]
\centering
\includegraphics[width=0.9\linewidth]{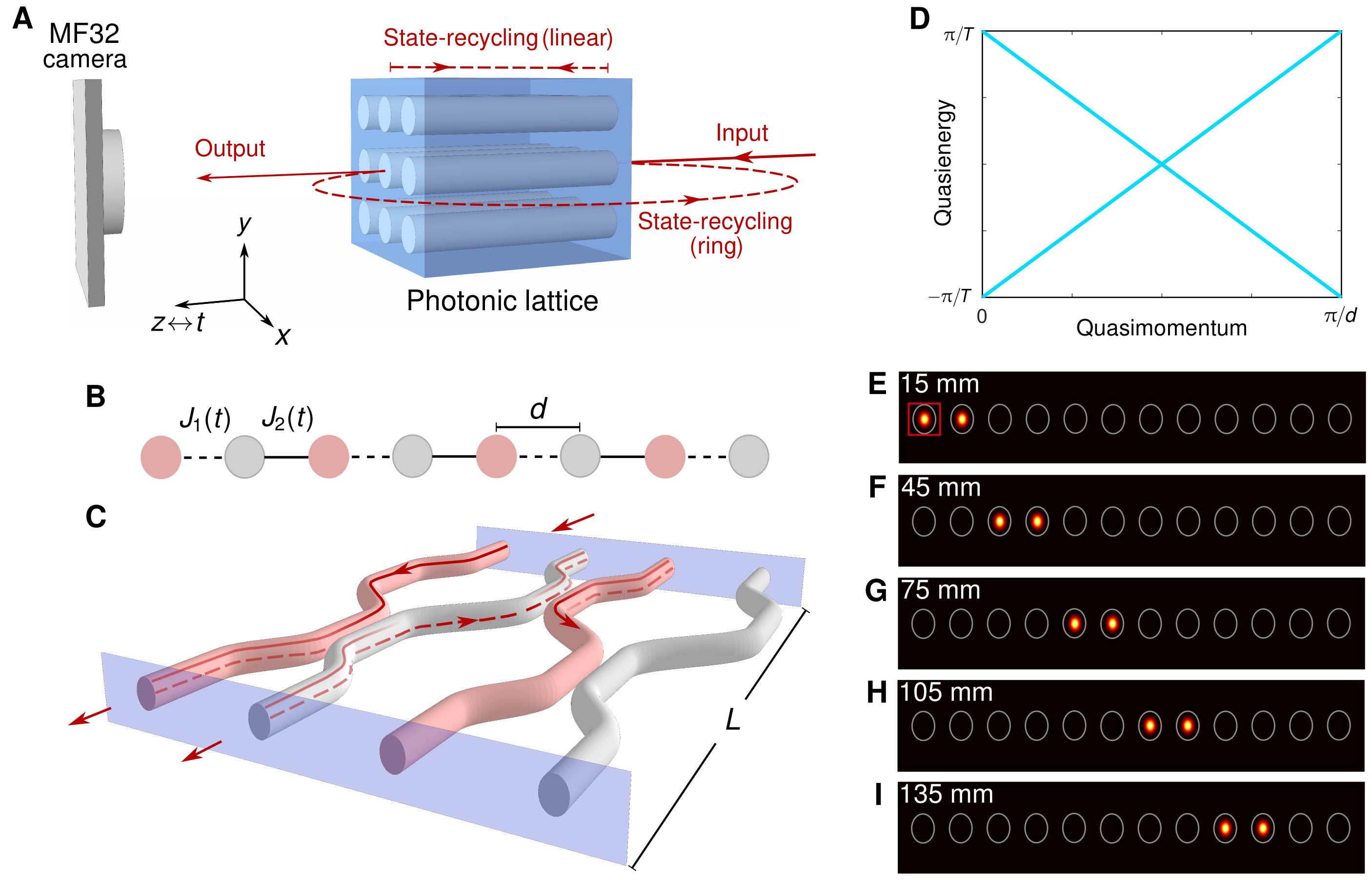}
\vspace{-0.cm} \caption{{\bf State-recycling techniques.} {\bf A,} Simplified sketch illustrating the experimental technique to detect `time' evolution in a photonic lattice through state-recycling. Both the ``linear" and ``ring" recycling schemes are illustrated. The propagation distance is the analogous time ($z\!\leftrightarrow\!t$).
{\bf B,} One-dimensional driven lattice with nearest-neighbour couplings $J_{1, 2}$, which are varying periodically in time. {\bf C,} Photonic implementation of the driven lattice in (B). Here, the state-recycling is performed using a linear cavity, the facets of which are  indicated by two parallel blue planes. {\bf D,} Floquet spectrum of the driven lattice in (B, C) consists of two linearly dispersive bands for the following driving protocol: $J_{1, 2}\!=\!0, \pi/T$ for $0\! \le \! t \le \! T/2$ and $J_{1, 2}\!=\! \pi/T, 0$ for $T/2\! \le \! t \! \le \!T$ where $T$ is the driving period. {\bf E-I,} Excitation of a linearly dispersive band. Experimentally observed intensity distributions at $t\!=\!(1/2+N)T$, $N\!=\! 0, 1, 2, 3$. The red square indicates the waveguide which was excited at the input. {\color{Seba}The effective propagation distances are indicated on each image.}
\label{fig_data}}
\end{figure*}

{\bf Results}\\
{\bf State-recycling schemes.} 
The key concept behind our state-recycling system is to place the photonic simulator inside an optical cavity, which allows the output state to be fed back into the simulator. As shown in Figure~\ref{fig_data}A, we consider two types of cavities. In the ``linear-cavity" scheme, light is simply reflected at both ends of the lattice, so that the effective time-evolution is dictated by an alternating sequence of time-evolution operators, $\hat U(t)=e^{-i \hat H_{2} T/2}e^{-i  \hat H_{1} T/2} \dots e^{-i \hat H_{2} T/2}e^{-i  \hat H_{1} T/2}$, where the Hamiltonians $\hat H_{1,2}$ are related by a time-reversal operation and $T$ is the period associated with each round trip. This linear cavity is most suitable to study the long-time dynamics associated with a specific engineered Hamiltonian $\hat H$, whenever the latter is time-reversal symmetric, $\hat H\!=\!\hat H_{1}\!=\!\hat H_{2}$. In the ``ring-cavity" scheme, the output state is recycled and re-injected directly into the input [Figure~\ref{fig_data}A]. This ring scheme is thus suitable to simulate Hamiltonians without time-reversal-symmetry, such as those associated with the quantum Hall effect and (Floquet) Chern insulators~\cite{Lu2014,Khanikaev2017}. 
By launching optical pulse trains at the input of a photonic lattice, we are  able to use an advanced single-photon sensitive detector array to perform independent time-correlated single-photon counting for each mode of the simulator, and thus to observe the time-evolution of the light field in a quasi-real-time manner. The key technology at the heart of our scheme is therefore the single-photon sensitive detector array itself, which in our case consists of a $32\!\times\!32$ square array of silicon based SPADs manufactured using complementary metal oxide semiconductor (CMOS) technology. Each SPAD has a $\approx 6\ \mu$m diameter photosensitive area and the pixel pitch is $50\ \mu$m. The photon detection efficiency of the SPADs is maximum at a wavelength of about $500$~nm, but is still single-photon sensitive up to about $1000$~nm. Each individual pixel can acquire time information with a resolution of $53$~ps for $10$ bits (i.e.~$54$~ns) temporal range. This type of detector array was recently used for a variety of multiplexed single-photon counting~\cite{Guerrieri2010} applications, including light-in-flight imaging~\cite{Gariepy2015} {\color{Seba}and multiplexed single-mode single-photon-sensitive wavelength-to-time mapping~\cite{Chandrasekharan2017}.}

{\bf 1D Dirac fermions.} 
We first apply the linear-cavity scheme to a periodically-driven lattice 
which emulates pure one-dimensional (1D) Dirac fermions~\cite{Budich_helical, dreisow2013, Bellec_2017}. In this model, the amplitudes of nearest-neighbour couplings are staggered and modulated in a periodic manner according to a two-step sequence [Figure~\ref{fig_data}B]: for the first half a period ($0\! \le \! t\! \le \!T/2$), neighbouring couplings are $J_1\!=\!0$ and $J_2\!=\!\pi/T$, while for the remaining half ($T/2\! \le t \! \le \! T$), these couplings are $J_1\!=\!\pi/T$ and $J_2\!=\!0$. In this configuration, the effective Hamiltonian associated with stroboscopic motion~\cite{Goldman2014periodically} takes the form of a Dirac Hamiltonian, $\hat H_{\text{eff}}\!=\! v_D k \hat \sigma_{z}$, where $k$ is the crystal momentum, $\hat \sigma_{z}$ is a Pauli matrix describing the lattice pseudo-spin, with the effective ``speed of light" $v_D\!=\!2 d/T$ where $d$ is the lattice spacing (see Supplementary Note 1). Accordingly, the Floquet spectrum consists of two linearly dispersive bands, Figure~\ref{fig_data}D, indicating that a single-band excitation is expected to travel along the lattice without any diffraction.

To implement this driving protocol, a photonic lattice of $24$ sites was fabricated inside a 15-mm-long borosilicate substrate using ultrafast laser inscription~\cite{Davis1996} (see Methods). Each waveguide pair was synchronously curved to spatially and dynamically turn on/off any particular bond [Figure~\ref{fig_data}C], hence generating the desired effective couplings~\cite{Mukherjee2017}. Both facets of the substrate were polished and silver-coated (with $\approx 90\%$ reflectivity) so as to form the above-mentioned linear cavity, and the optical mode of each waveguide was imaged from the output of the lattice onto individual SPADs of the Megaframe (MF32)~\cite{Richardson2009}, see Supplementary Figure~1. In our setup, the actual photonic lattice only describes half of the complete driving sequence described above (i.e.~$L\!\equiv\!T/2$). However, thanks to the linear cavity, and due to the time-reversal nature of the underlying effective Hamiltonian, our photonic lattice enables us to launch an initial state at the effective time $t\!=\!0$ and to detect it at stroboscopic times  $t\!=\!(1/2+N)T$, where $N$ is a positive integer. 
 
\begin{figure*}[]
\centering
\includegraphics[width=0.95\linewidth]{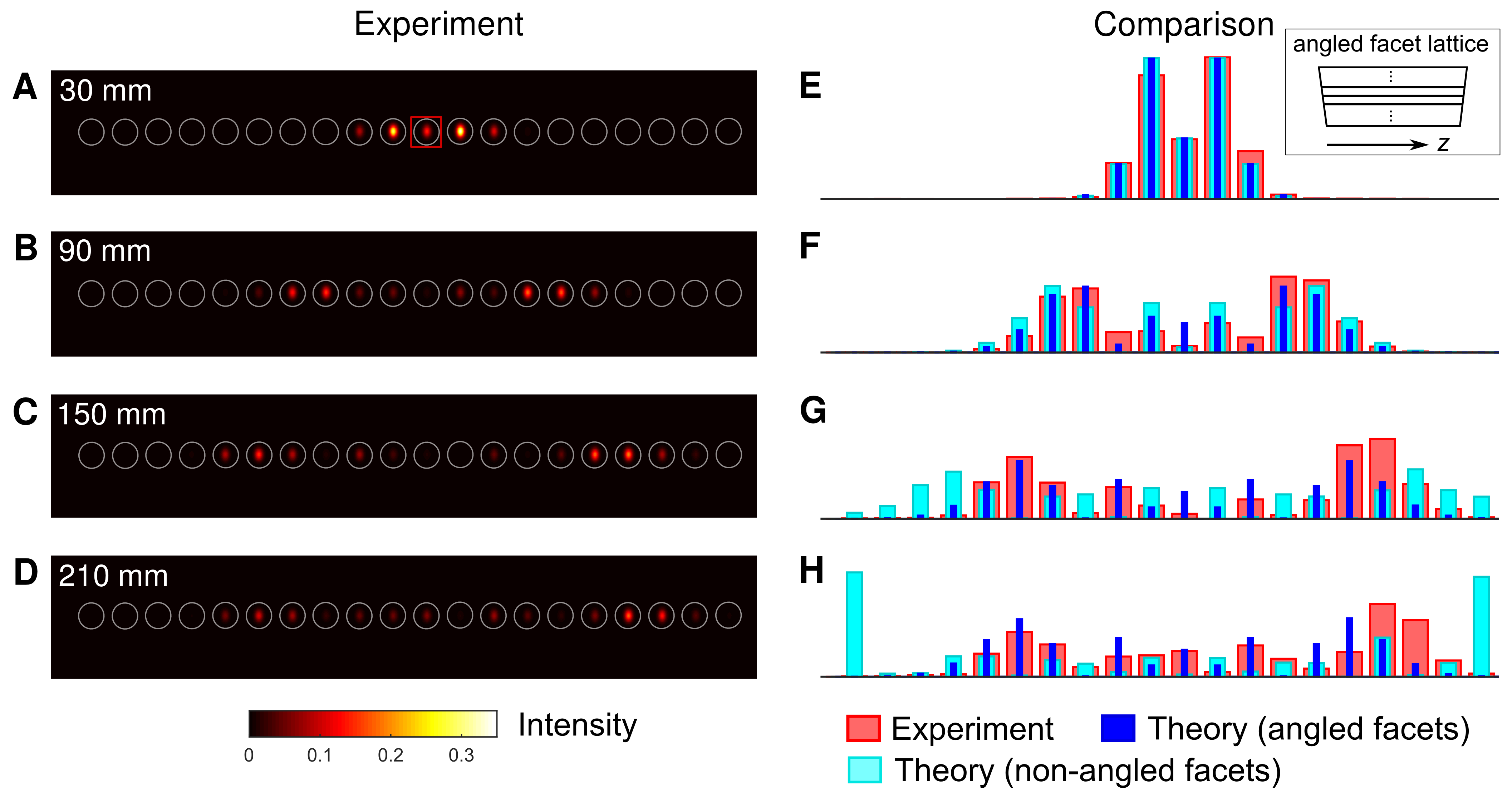}
\vspace{-0.cm} \caption{ {\color{Seba}
{\bf Discrete diffraction in the presence of a synthetic electric field.} {\bf A-D,} Quasi-real-time evolution of light intensity in a 1D straight photonic lattice consisting of twenty coupled single-mode waveguides. The small facet angles 
of the substrate (inset in E) cause a linear phase shift along the lattice at each facet, which effectively produces a time-periodic (pulsed) synthetic electric field. A-D show the output intensity distributions that were experimentally obtained after different effective propagation distances indicated on each image.
Light was launched at a single waveguide indicated by the red square. {\bf E-H,} Comparison between the experimental observations in (A-D) and the associated numerical results. The vertical axis is the normalized optical intensity and the horizontal axis is the waveguide number. The measured intensity patterns (red bars) agree with the numerical simulations (blue) upon adding the effects of a pulsed (synthetic) electric field. The dynamics corresponds to approximately half a period of a Bloch oscillation, which is produced by the synthetic electric field. For comparison, we show the numerical results when omitting the pulsed electric field (cyan bars), where the Bloch oscillation is absent. }
\label{fig_discrete}}
\end{figure*}

In the experiment, we launched $873\!\pm\!3$~nm pulses at the edge of the lattice, and we measured the time-correlated photon counts up to four and a half round trips. This effectively corresponds to exploring
nine times the physical length of the lattice (i.e.~$9L\!=\!135$ mm), see Figure~\ref{fig_data}E-I. As shown, the data recorded by individual pixels provide spatial as well as temporal intensity distributions, see Supplementary Figure~2. 
The number of round trips that can be observed in our current experimental system depends on the quality factor of the cavity, which is primarily determined by the waveguide losses ($\sim\!5$~dB per each round trip for this experiment). In other words, the temporal data can be accessed until the detected photon count is comparable to the noise level. Exploiting suitable detectors and performing the experiment at a longer wavelength (e.g.~near $1550$~nm, where comparatively low-loss waveguides can be fabricated) will allow one to access many more round trips.

{\color{Seba}
{\bf Discrete diffraction and synthetic electric fields.} 
A similar linear-cavity setup was used to investigate discrete diffraction in a 1D photonic lattice consisting of twenty coupled single-mode waveguides with $19 \ \mu$m waveguide-to-waveguide spacing and length $L\!=\!30$ mm. 
In this situation, the propagation of optical fields emulates the motion of a particle in a (single-band) tight-binding lattice~\cite{christodoulides2003discretizing}. 
Figures~\ref{fig_discrete}A-D show the intensity patterns that were experimentally obtained for four effective propagation distances, $L\!=\!30$ mm, $3L\!=\!90$ mm, $5L\!=\!150$ mm and $7L\!=\!210$ mm. 
Interestingly, as shown in Figures~\ref{fig_discrete}E-H, these intensity patterns (red bars) start to deviate from the ``naively expected" distributions (cyan) for effective propagation distances $z\!>\!3L$. This is due to the small angles located at the input and output facets ($\approx\!\pm 0.1^{\circ}$, respectively) of the substrate (see inset of Fig.~\ref{fig_discrete}E), which cause a linear variation of the optical phase along the lattice axis. 
These angles {\it effectively} produce a time-periodic (pulsed) electric field that acts on the particle along the lattice axis; see Supplementary Notes~3 and 4. This picture is validated in Figures~\ref{fig_discrete}E-H, which indicates that the experimental intensity patterns (red bars) agree well with our numerical simulations upon adding the effects of the pulsed electric field (blue bars). The resulting motion is found to correspond to approximately half the period of a Bloch oscillation; see Supplementary Note~4. We note that effects of the small facet angles were not detected in the previous experiment (Figure~\ref{fig_data}E-I) because the spatial extent of the analogous wavefunction was significantly smaller as compared to the current experiment (Figure~\ref{fig_discrete}).
}

\begin{figure*}[]
\centering
\includegraphics[width=1.00\linewidth]{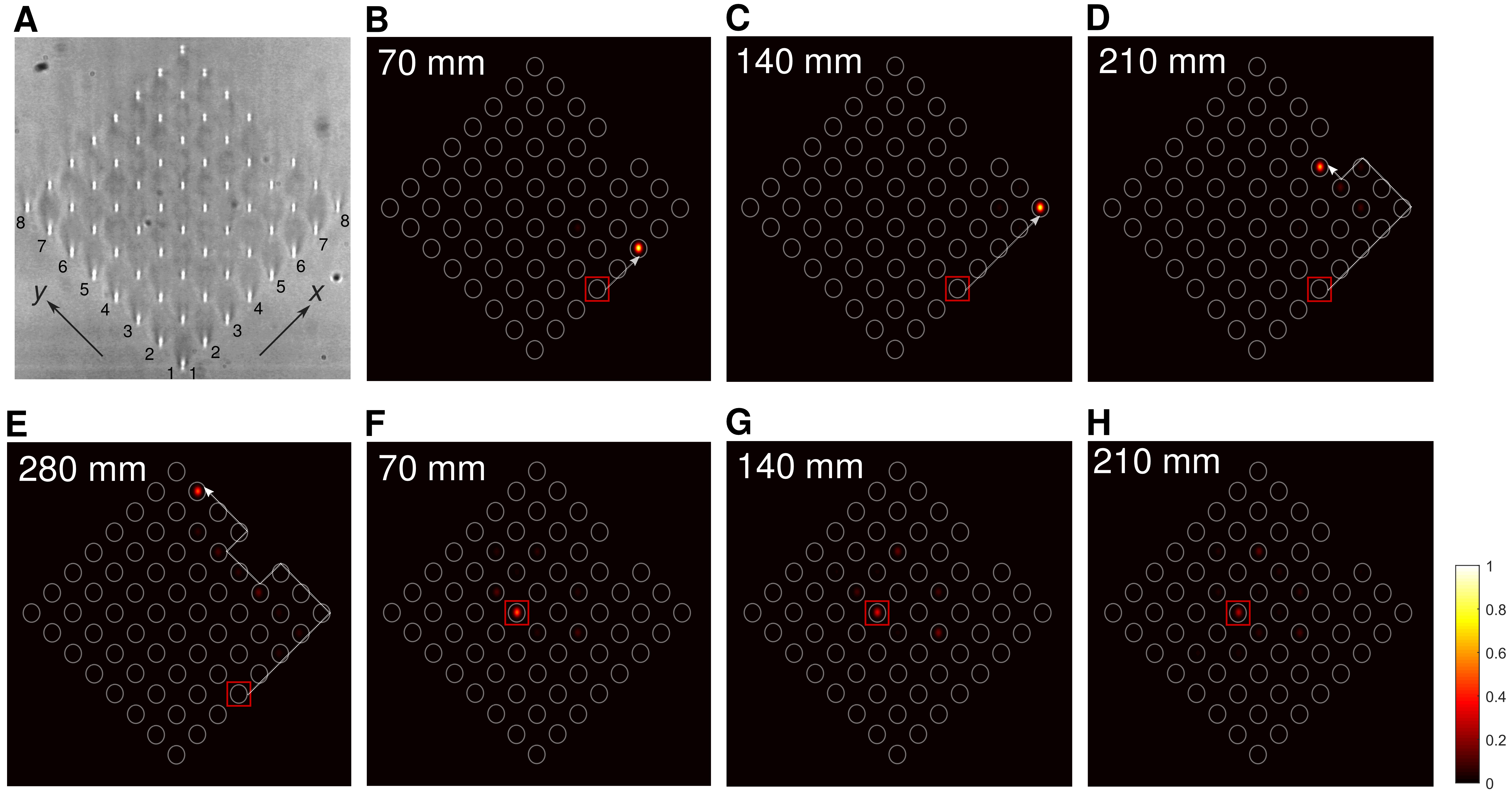}
\vspace{-0.cm} \caption{{\bf Quasi-real-time propagation of topological edge modes.} {\bf A,} White-light micrograph of the facet of the driven square lattice. 
{\bf B-E,} One-way (here, counter-clockwise) propagation of the edge modes for effective times, $t\!=\!2T, \ 4T, \ 6T$ and $8T$; {\color{Seba}here $2T\!\leftrightarrow\!L\!=\!70 \ $mm.}
The edge modes are excited with $\approx85\%$ efficiency by exciting the $(4, 1)$ site on the bottom-right edge (indicated by the red square). 
The edge modes are neither back-scattered by a corner nor by the defect [here, a missing waveguide at the $(8, 4)$ site]. {\bf F-H,} Time evolution of the bulk state after times, $t\!=\!2T, \ 4T$ and $6T$. The weakly dispersive bulk bands are equally excited by launching light at the $(4, 5)$ site. The delocalization of the state becomes evident after long detection times.
\label{fig_EDGE_BULK}}
\end{figure*}

{\bf Anomalous Floquet topological insulator.} 
Next, we demonstrate the operation of the ring-cavity state-recycling method. First of all, we verified that both the phase and intensity of an optical state are recycled in the ring cavity as required [see Methods and Supplementary Fig.~3]. Here, we discuss the application of the ring-type cavity to a two-dimensional periodically-driven system exhibiting non-trivial topology~\cite{Kitagawa2010,Rudner2013}. The model realised in our photonic setup was introduced in Ref.~\citen{Rudner2013} to demonstrate the existence of {\it anomalous} topological edge modes, which are topological states appearing in periodically-driven (Floquet) systems with no static-system counterparts. Formally, these robust propagating states are protected by a topological winding number, which, in contrast to the more conventional Chern number~\cite{Hasan2010}, takes into account the full-time dynamics of the time-modulated system~\cite{Rudner2013}. Such anomalous topological edge modes were experimentally demonstrated in photonics~\cite{Gao2016, Mukherjee2017, Maczewsky2017}. Here, we describe how our state-recycling technique can be applied to such intriguing states of matter, and exploit to reveal the quasi-real-time imaging of the corresponding chiral topological edge modes. We point out that the simulated system explicitly breaks time-reversal symmetry, and therefore it cannot be explored using the simpler linear-cavity scheme.

The Floquet model of Refs.~\citen{Rudner2013,Mukherjee2017, Maczewsky2017} consists of a driven square lattice with four distinct nearest-neighbour couplings $(J_{1-4})$, which are varied in a circulating and time-periodic manner over the entire lattice, see Supplementary Note~2. When the driving period $(T)$ is split into four equal steps, and upon the resonance condition $J_{1-4}\!=\!2\pi/T$, chiral propagating anomalous edge modes are found to coexist with a perfectly localized bulk [see Supplementary Figure~4A]. However, in this case, the Floquet bulk bands are degenerate at zero quasienergy, and an arbitrarily small deviation in the values of the parameters can potentially drive the system out from the anomalous regime~\cite{Mukherjee2017}. 
To avoid possible ambiguity, and also for the sake of experimental practicability, we designed a slightly different model with $J_{1}\!=\!0$ and $J_{2,3,4}\!=\!2\pi/T$.  
In this situation, maximally gapped bulk bands appear with zero Chern number [see Supplementary Figure~4B]., while the winding numbers associated with both the energy gaps (centred on $0$ and $\pi/T$) are non-trivial and equal to one.

Such a photonic lattice of $63$ sites [see Figure~\ref{fig_EDGE_BULK}A] was fabricated, corresponding to only two driving periods (i.e.~$2T\leftrightarrow\!L\!=\!70 \ $mm).
Initially, all the waveguides are uncoupled at the wavelength range of interest [nearest-neighbour spacing is $40 \ \mu$m]. Similar to the previous experiment (Fig.~1C), each waveguide pair is synchronously curved to spatially and/or dynamically turn on and off any particular bond~\cite{Mukherjee2017}. To investigate the effect of a defect, one waveguide on the top-right edge [$(8, 4)$ lattice site] was not fabricated. For the proposed experimental parameters, the edge modes, as well as the bulk states, can be efficiently excited by launching light into one optical waveguide.

In the experiment, we launch $780\pm5$~nm pulses of light at a desired site of the photonic lattice, which is placed inside the ring-type cavity, see supplementary Figure~1. The output facet of the lattice is imaged onto the input of the lattice with unit magnification. For precise imaging, the input facet of the lattice is imaged on a CCD camera to observe the lattice sites, input state, and the output state after the first pass. Similar to the previous experiments,  the optical mode of each waveguide from the output of the lattice is imaged onto the SPAD array. 

The edge modes are excited with $\approx85\%$ efficiency by launching light at the $(4, 1)$ site on the bottom-right edge. Quasi-real-time chiral propagation of the edge modes is presented in Figure~\ref{fig_EDGE_BULK}B-E for four consecutive round trips (i.e.~at analogous time $2T$, $4T$, $6T$ and $8T$). It should be noted that the group velocity of the edge modes along the top-right edge is twice that along the bottom-right edge~\cite{Mukherjee2017}. For the proposed parameters, the non-dispersive bulk bands can be excited equally by launching light at a waveguide in the bulk [e.g.~site (4, 5) in Figure~\ref{fig_EDGE_BULK}F, H]. In this situation, the input state is expected to cycle back to its initial position after two complete driving periods.
Although the bulk bands are expected to be dispersionless for the proposed parameters, a small deviation from the desired parameters value makes both bands weakly dispersive~\cite{Mukherjee2017} without altering the topological characteristics of the system. The effect of this weak dispersion of the bulk bands is hard to detect after first round trip, see Figure~\ref{fig_EDGE_BULK}F. However, the weak delocalisation of the bulk state becomes evident when exploiting the many round trips offered by the cavity [Figures~\ref{fig_EDGE_BULK}G, H], which demonstrates the capability of our state-recycling technique to detect the slow dynamics of such weakly-dispersive states.

\newpage 

{\bf Discussion}\\
In conclusion, we have proposed and experimentally demonstrated a state-recycling technique based on time-correlated single-photon counting imaging, which enables us to measure the long-time dynamics of an input optical state propagating in an engineered photonic lattice. Importantly, this method introduces the possibility of detecting effective dynamics in a quasi-real-time (stroboscopic) manner. It offers a novel dimension to photonic lattices, for which the final detection time was until now set by the length of the photonic lattice.  Furthermore, the ring-cavity method, in which the state is re-injected into the lattice in a controllable manner, offers a unique opportunity to design feedback mechanisms, i.e.~a hybrid analogue-digital simulator. For instance, modifying the state after each round trip, according to some well-defined unitary operators, could be used as a simple protocol to design quantum walks~\cite{Kitagawa_walk, Cardano2017}, or could be suitably combined with another Floquet-engineering protocol. As we discussed, such stroboscopic operations can be used to simulate the effects of external effective fields (e.g.~forces), which could allow one to perform transport experiments~\cite{Hasan2010, Heiblum_edge} within the photonic lattice over long timescales. Similar operations could be used to design dynamical (density-dependent) gauge fields~\cite{Edmonds2013}, to imprint effective (mean-field) interactions~\cite{Lumer_interactions}, or to engineer space- and time-dependent losses in the photonic lattice~\cite{Rudner_Levitov}.

\begin{methods}
{\bf Fabrication.} The photonic lattices were fabricated inside borosilicate substrates (Corning Eagle$^{2000}$) using ultrafast laser inscription~\cite{Davis1996}.
The substrate was translated at $8$ mm/s once through the focus of sub-picosecond laser pulses ($350$~fs, $500$~kHz, $1030$~nm) to fabricate each waveguide. The pulse energy of the laser was optimised to realise tightly confined single mode waveguides for a desired wavelength. 

For emulating Dirac fermions and the anomalous Floquet topological phase,
we used synchronously bent waveguide pairs to turn the bonds on and off; see Ref.~\citen{Mukherjee2017}. Initially (i.e.~at $z\!=\!0$), all the waveguides in the lattices are well separated such that the inter-waveguide couplings are insignificant. To turn on coupling between any desired waveguide pair, we reduce the inter-waveguide separation by synchronously bending the waveguide axes. The waveguides then propagate parallel to each other for a certain length and finally separate in a reverse manner. The coupling between two such bent waveguides is equivalent to an effective tight-binding coupling between two straight neighbouring waveguides. The effective bond strength depends on the geometry of the bending profile. For the precise control of the bond strength, we tune the wavelength of the excitation light. 

{\bf Linear and ring cavity schemes.} The experimental setup for the time-resolved state-recycling is shown in supplementary  Figure~1. 
In the experiment, light at $39$~MHz pulse repetition rate and a desired wavelength (determined by a bandpass filter, F) is filtered from a broad-band supercontinuum source (NKT Photonics). The beam splitter, BS$_1$, reflects $\approx10 \%$ of this light which enters the ring cavity (formed by M$_{2-5}$).  For precise imaging, the input facet of the lattice is imaged on a CCD camera to observe the lattice sites, input state, and the output state after the first pass. The optical mode of each waveguide at the output of the photonic lattice is imaged onto individual SPADs of the Megaframe (MF32); [similar devices are now supplied commercially by Photon Force Ltd]. For the experiments shown in Figure~\ref{fig_data}C and Figure~\ref{fig_discrete}, where the state-recycling is performed using a linear cavity, the aforementioned lattice is replaced by one with silver-coated facets (shown in the green-dotted inset) and BS$_3$ is replaced by a mirror to reflect the output state to the MF32. Supplementary Figure~2A 
shows an optical micrograph of the MF32 camera with the $32\!\times\!32$ SPAD array.

As mentioned in the main text, time-correlated single photon counting by the silicon-based SPAD array provides access to both spatial and temporal information. The data processing method for the driven 1D lattice (presented in Figure~\ref{fig_data}E-I) is briefly summarized in supplementary Figure~2B-C. 
Supplementary Figure~2B 
shows the spatial information i.e.~intensity distribution summed over four and a half round trips and the normalized temporal information for some specific pixels (indicated by the arrows) are presented in supplementary Figure~2C. 
It should be noted that the temporal separation ($\tau_s$) between two consecutive peaks is determined by the length of the cavity. For this particular experiment, we used a $15$-mm-long cavity, hence,  $\tau_s\!\sim\!150\ $ps. In this situation, it is expected that the temporal separation $\tau_s$ will be $3$ bins (i.e.~time-steps), as observed in supplementary Figure~2C. 
Figure~\ref{fig_data}E-I 
show the evolution of the intensity distribution along the lattice which was obtained considering the peak intensities of the recorded signals [supplementary Figure~2C]. 

{\color{Seba}
{\bf  Intensity and phase recycling in the ring cavity.} To demonstrate that an optical state (both phase and intensity) is recycled in the ring cavity, we perform the following experiment. A directional coupler, formed by two evanescently coupled straight optical waveguides was fabricated and placed inside the ring cavity. The coupling strength for this device were measured to be $J\!=\!0.046$ and $0.038$ per mm at $780$~nm and $750$~nm wavelengths, respectively. First, we launched optical pulse trains at $780$~nm into waveguide-1 and measured output intensities at both waveguides in a time-resolved manner. The blue and red solid lines in Supplementary Figure~3A 
shows the expected variation of light intensities, $I_1$ and $I_2$, as a function of the dimensionless parameter, $Jz$. The red and blue squares indicate the measured intensities at $780$~nm wavelength for four consecutive round trips.  Next, we used pulse trains at $750$~nm to reduce the coupling strength and the corresponding intensities are indicated by red and blue circles in Supplementary Figure~3A. 
The distributions of light intensity measured after each round trip can only be observed if both the phase an amplitude of the state is preserved during the state-recycling process, confirming this to be the case.

In addition, a time-resolved interference experiment was performed at $750$~nm.  Supplementary Figures~3B-E 
show interference fringes (between the two output modes) detected by the MF32 for four consecutive round trips. The fringes are rotated at $45^{\circ}$ because the waveguides were oriented at that angle with respect to the vertical axis. The $\pi$ phase shift observed in Supplementary Figures~3D and E 
compared to B and C is a well-known characteristic of a directional coupler -- after the full transfer of light, i.e.~$Jz\!>\!\pi/2$, the relative phase between the optical modes of the waveguides exhibit a phase shift of $\pi$. 
}
\end{methods}

\begin{addendum}
\item[Data availability] Raw experimental data will be made available through Heriot-Watt University PURE research data management system.
\end{addendum}

{\bf References} \vspace{-4ex}



\begin{addendum}
\item This work was funded as part of the UK Quantum Technology Hub for Quantum Communications Technologies - EPSRC grant no. EP/M013472/1), and by the UK Science and Technology Facilities Council (STFC) - STFC grant no. ST/N000625/1. N. G. is financially supported by the FRS-FNRS (Belgium) and the ERC TopoCold Starting Grant. P. \"O. acknowledges support from EPSRC grant no. EP/M024636/1.
We thank R. K. Henderson for providing the SPAD array used in this work. We also thank E. Andersson, M. Hartmann, H. M. Price, A. Spracklen and M. Valiente for helpful discussions.

\end{addendum}

\newpage

\renewcommand{\figurename}{{\bf Supplementary Figure}}

\newcommand{\beginsupplement}{%
        \setcounter{equation}{0}
        \renewcommand{\theequation}{\arabic{equation}}%
        \setcounter{figure}{0}
        \renewcommand{\thefigure}{{\bf \arabic{figure}}}%
     }
\beginsupplement

\section*{\centering{\large{Supplementary Information}}}

\bigskip 

\begin{figure}[h!]
\centering
\includegraphics[width=0.9\linewidth]{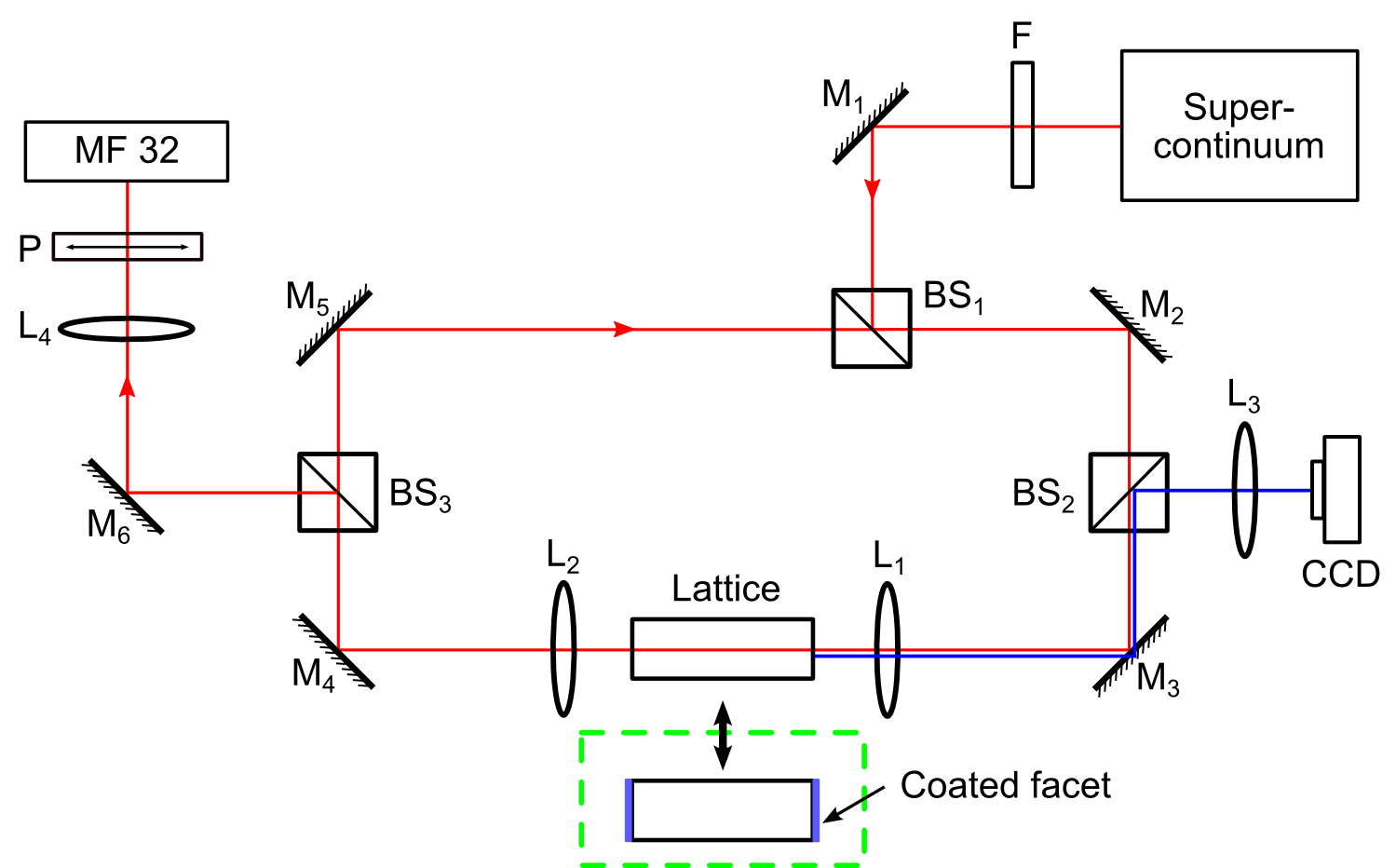}
\vspace{-0.cm} \caption{{\bf Experimental setup.} Here, L$_{1,5}$ are convex lenses, M$_{1-6}$ are silver-coated mirrors, F is a bandpass filter at $780\!\pm\!5$ nm wavelength, BS$_{1-3}$ are beam splitters and P is a polarizer. The input facet of the lattice is imaged on a CCD camera to observe the lattice sites, input state, and the output state after the first pass.
To perform the state recycling using a linear cavity, the lattice is replaced by one with silver-coated facets (shown in the green-dotted inset), BS$_3$ is replaced by a mirror to reflect the output state to the SPAD array and F is replaced by another bandpass filter at $873\!\pm\!3$ wavelength.
\label{fig_ring_cavity}}
\end{figure}

\begin{figure}[t]
\centering
\includegraphics[width=1\linewidth]{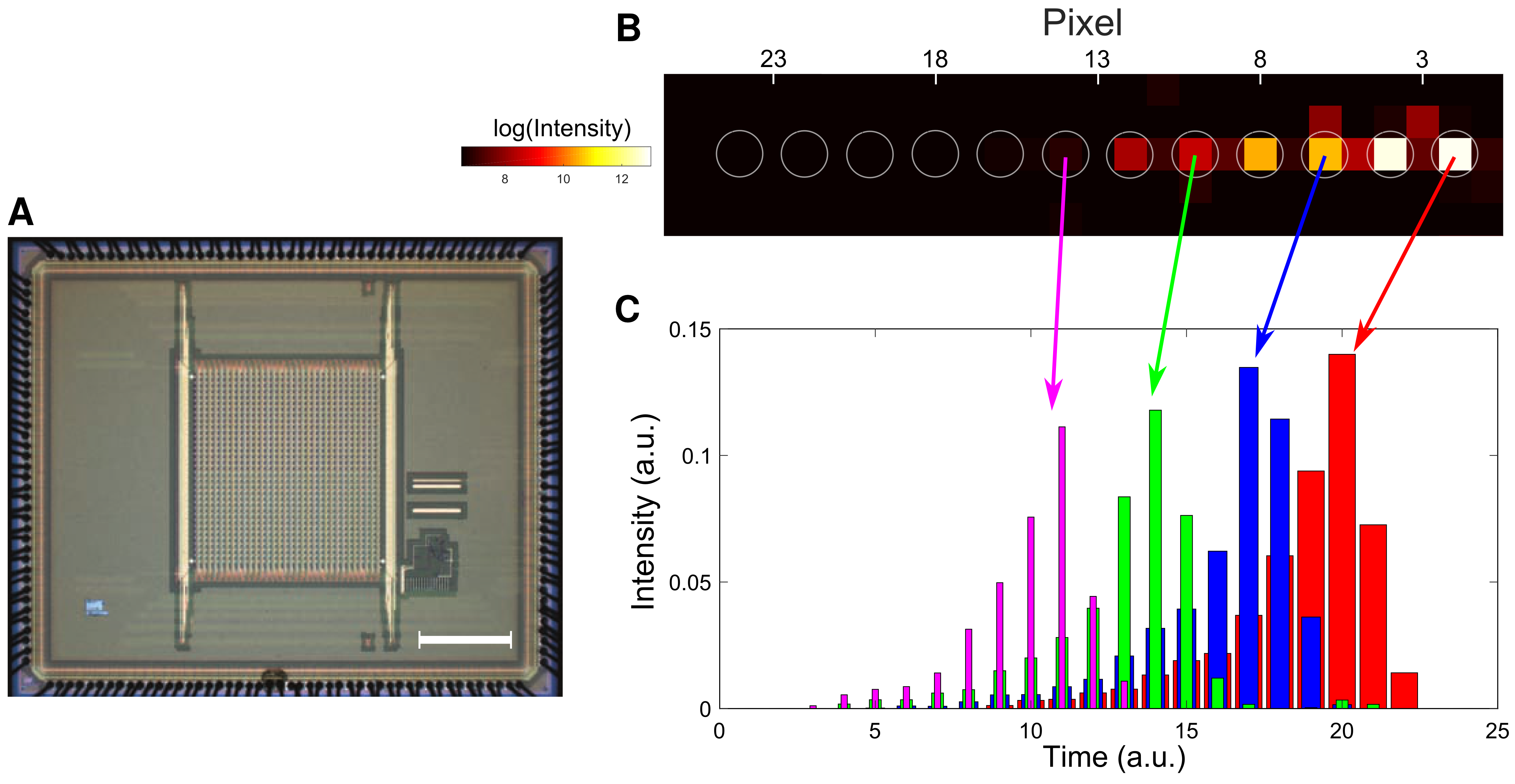}
\vspace{-0.cm} \caption{{\bf Data processing.} {\bf A,} Optical micrograph of the  Megaframe (MF32) camera with the  $32\!\times\!32$ SPAD array, scale-bar: $800\ \mu$m. {\bf B,} Spatial information i.e.~intensity distribution summed over four and a half round trips for the driven one-dimensional lattice presented in Figure 1 in the main text. To reduce spatial overlap of light intensity from two consecutive lattice sites, optical modes were imaged onto alternative pixels of the MF32 (represented by white circles). Each pixel contains temporal information. {\bf C,} Temporal intensity profiles for four pixels indicated by the colored arrows. These peaks are normalized such that the total detected optical power for each round trip is unity. Here, the time-step (horizontal axis) is $\sim\!53$ ps.
\label{data}}
\end{figure}

\begin{figure}[]
\centering
\includegraphics[width=0.9\linewidth]{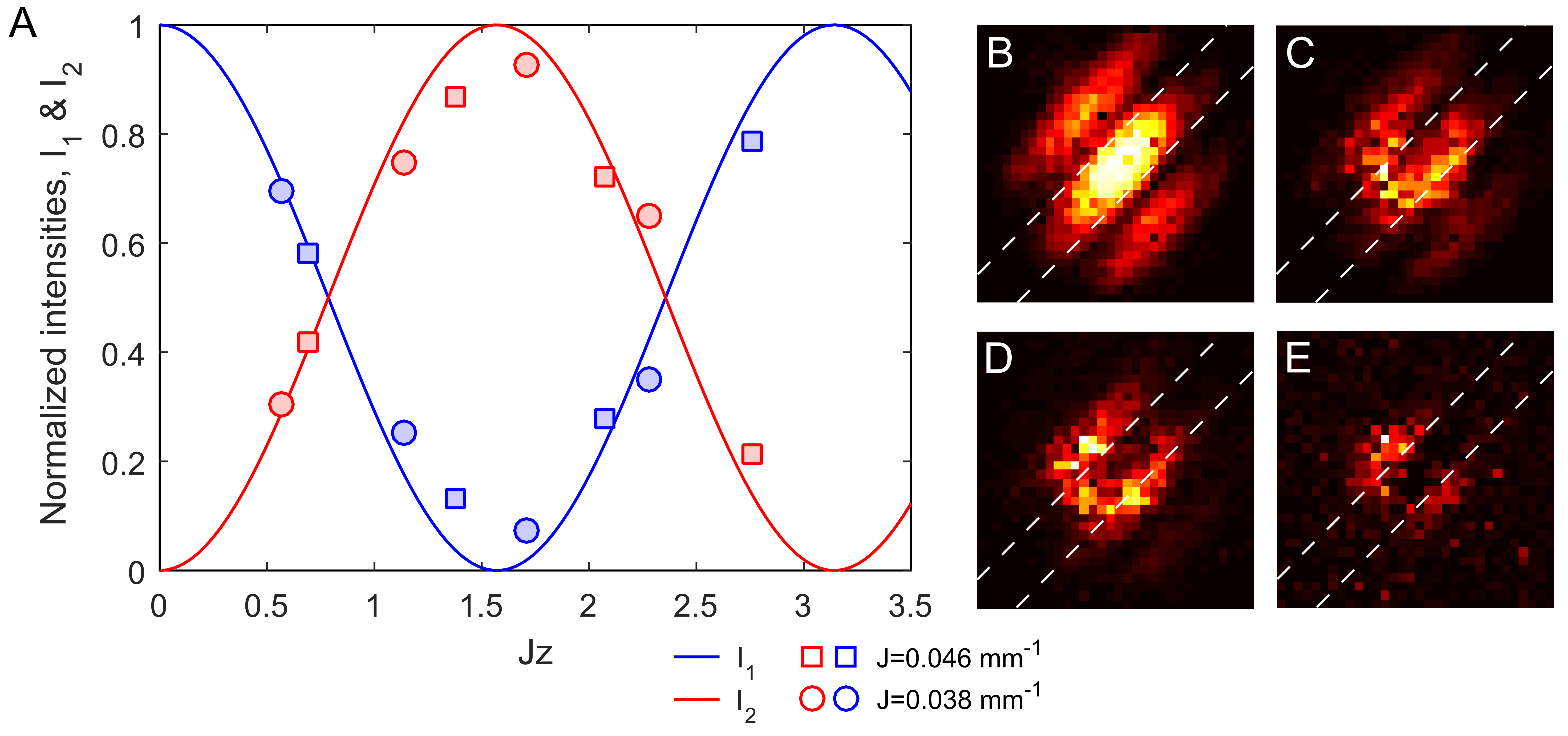}
\vspace{-0.cm} \caption{
{\bf Intensity and phase recycling in the ring cavity. A,} A directional coupler formed by two evanescently coupled waveguides was placed inside the ring cavity, light was initially launched at waveguide-1 and the evolution of optical fields was measured in a time-resolved manner. The solid lines indicate the expected variation of light intensity at waveguide-1 (blue) and 2 (red) as a function of $Jz$. Two sets of experiments were performed using optical pulse trains at $780$~nm (square) and $750$~nm wavelength (circle) which corresponds to two sets of tunnelling strengths, $J\!=\!0.046$ and $0.038$ per mm, respectively. {\bf B-E,} Time-resolved interference experiment at $750$~nm wavelength. B-E show interference fringes (at four consecutive round trips)
generated by allowing the modes at the output of the coupler to interfere on the SPAD array in the far-field. The fringes are rotated at $45^{\circ}$ because the
coupling axis between the waveguides was orientated
at that angle with respect to the vertical axis. The dashed lines are guides to the eye. The $\pi$ phase shift observed in D and E compared to B and C is a characteristic of a directional coupler -- after the full transfer of light, i.e.~$Jz\!>\!\pi/2$, the relative phase between the optical modes of the waveguides exhibit a phase shift of $\pi$. These simple experiments prove that both phase and intensity of optical fields is recycled in the ring cavity scheme.
\label{coupler}}
\end{figure}

\begin{figure}[t]
\centering
\includegraphics[width=1\linewidth]{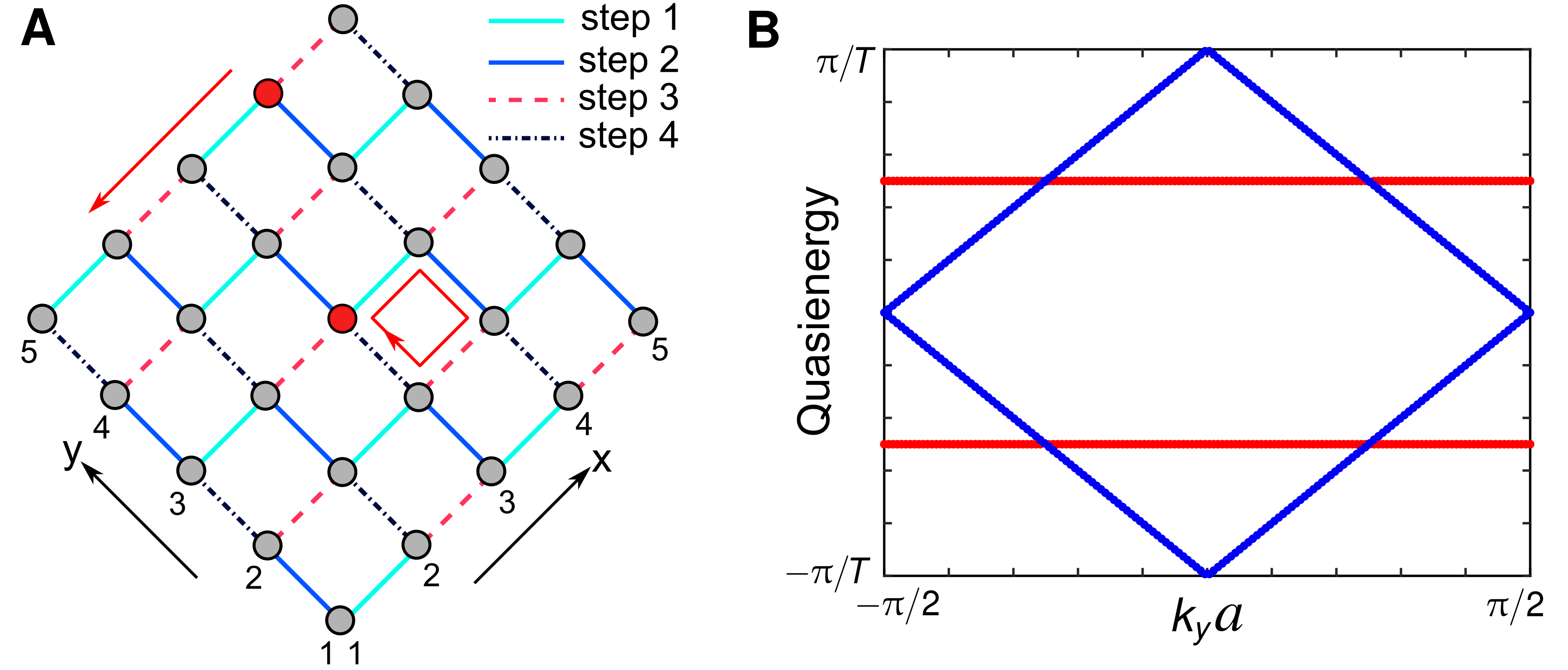}
\vspace{-0.cm} \caption{{\bf Driving protocol and band structure of a slowly-driven photonic lattice.} {\bf A,} A slowly-driven square lattice with nearest-neighbour couplings ($J_{1-4}$ corresponding to step {1-4}), which are varied in a spatially homogeneous and time-periodic manner. For $J_{1-4}T/4\!=\!\pi/2$, anomalous edge modes coexist with a localised bulk. In this case, the Floquet bulk bands are degenerate at zero quasienergy. {\bf B,} Floquet quasienergy spectrum for  $J_{1}T/4\!=\!0$ and $J_{2,3,4}T/4\!=\! \pi/2$. Here, the bulk bands (shown in red) with zero Chern number are gapped while the winding numbers associated with both the energy gaps (centred on $0$ and $\pi/T$) are one.
\label{fig_driving_protocol}}
\end{figure}

\begin{figure}[]
\centering
\includegraphics[width=1.0\linewidth]{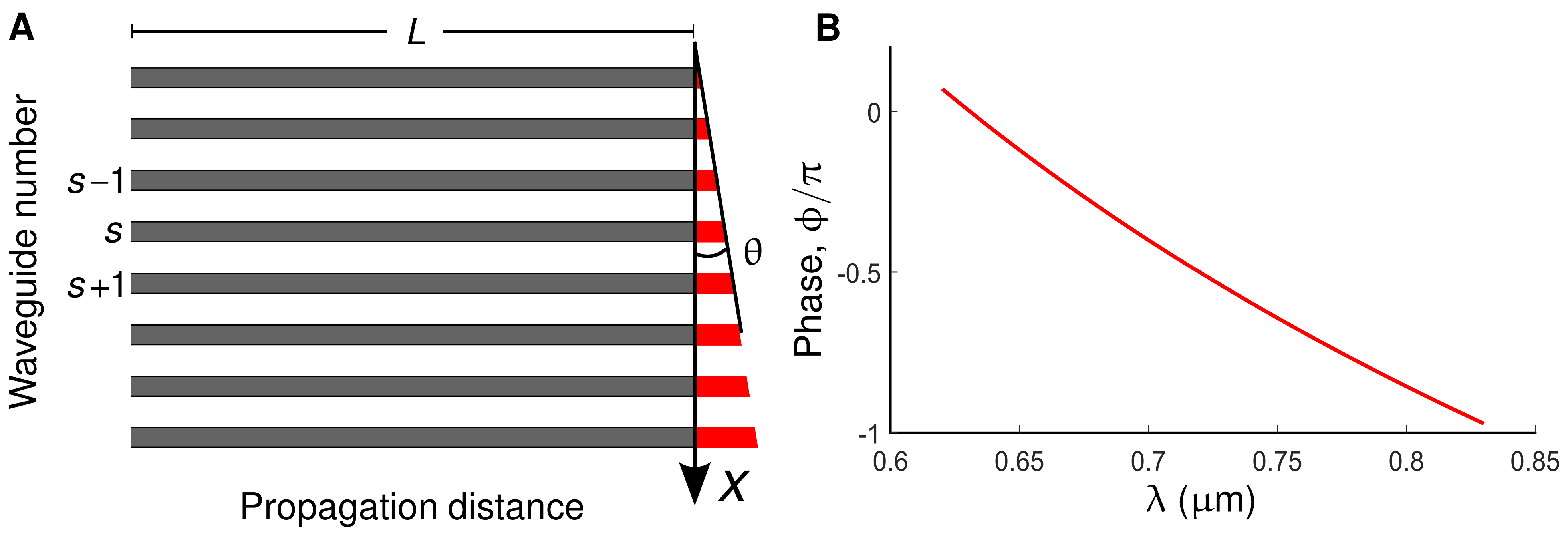}
\vspace{-0.cm} \caption{{\bf Floquet engineering of an electric field.} 
{\bf A,} The output facet of the photonic lattice is polished at a small angle $\theta$ with respect to $x$ axis, to realize an instantaneous electric field pulse which exists for an infinitesimally short duration of time. {\bf B,} For a specific value of $\theta\!=\!2^{\circ}$, the inter-waveguide phase shift ($\phi$) can be tuned by varying the wavelength of incident light, $\lambda$. 
\label{angled-lattice}}
\end{figure}
\begin{figure}[]
\centering
\includegraphics[width=1\linewidth]{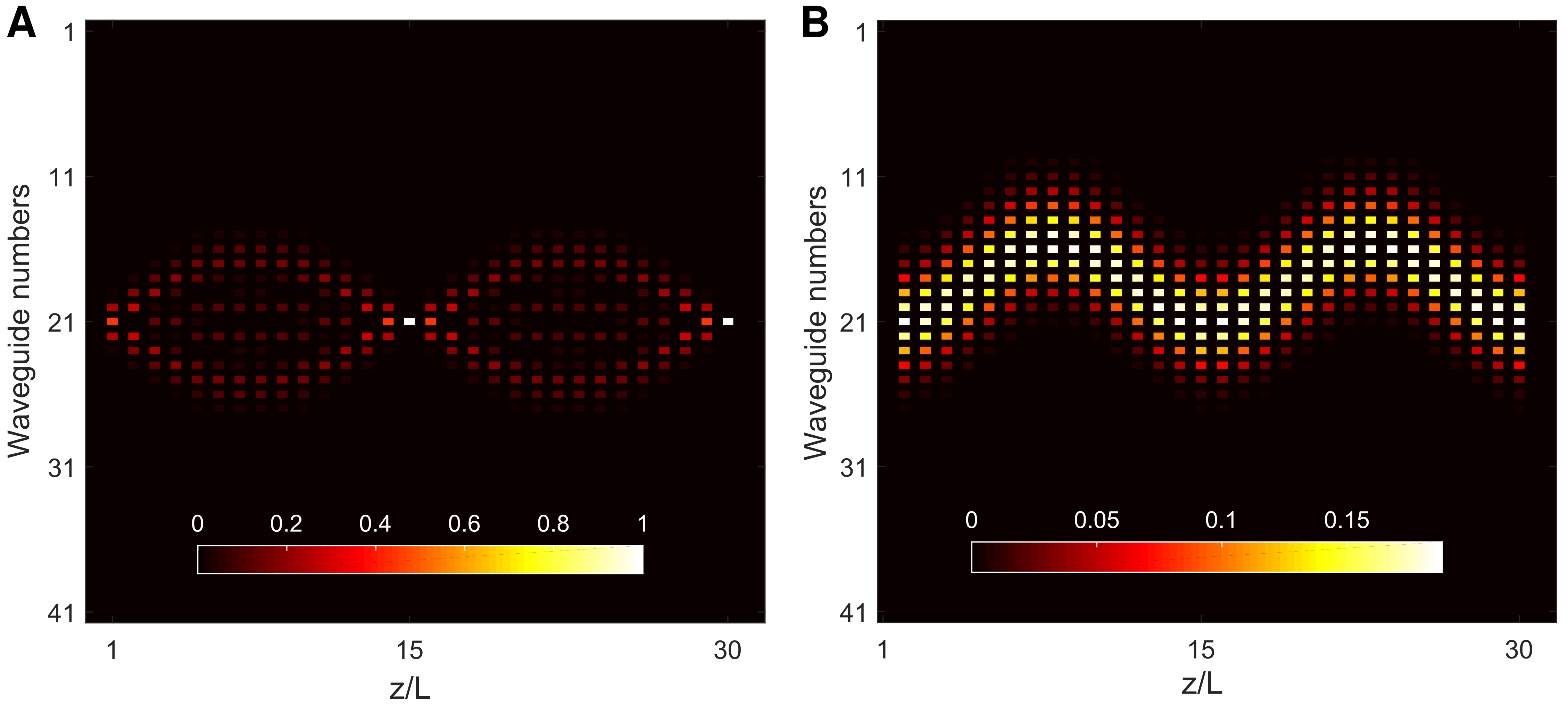}
\vspace{-0.cm} \caption{{\bf Time-periodic pulsed electric field along the lattice axis and Bloch oscillations.} 
A finite photonic lattice with $41$ waveguides and $\phi\!=\!\pi/7.5$ is considered. Here $J\!=\!0.02 \ $mm$^{-1}$ and $L\!=\!30 \ $mm. {\bf A,} Numerically calculated evolution of optical intensity when only the $21$-st waveguide is excited initially; here a breathing motion of the intensity pattern is observed. {\bf B,} In this case a broad initial state (in real space) is considered and the oscillation of the wave-packet's center of mass is observed.
\label{BO}}
\end{figure}

\begin{figure}[]
\centering
\includegraphics[width=0.8\linewidth]{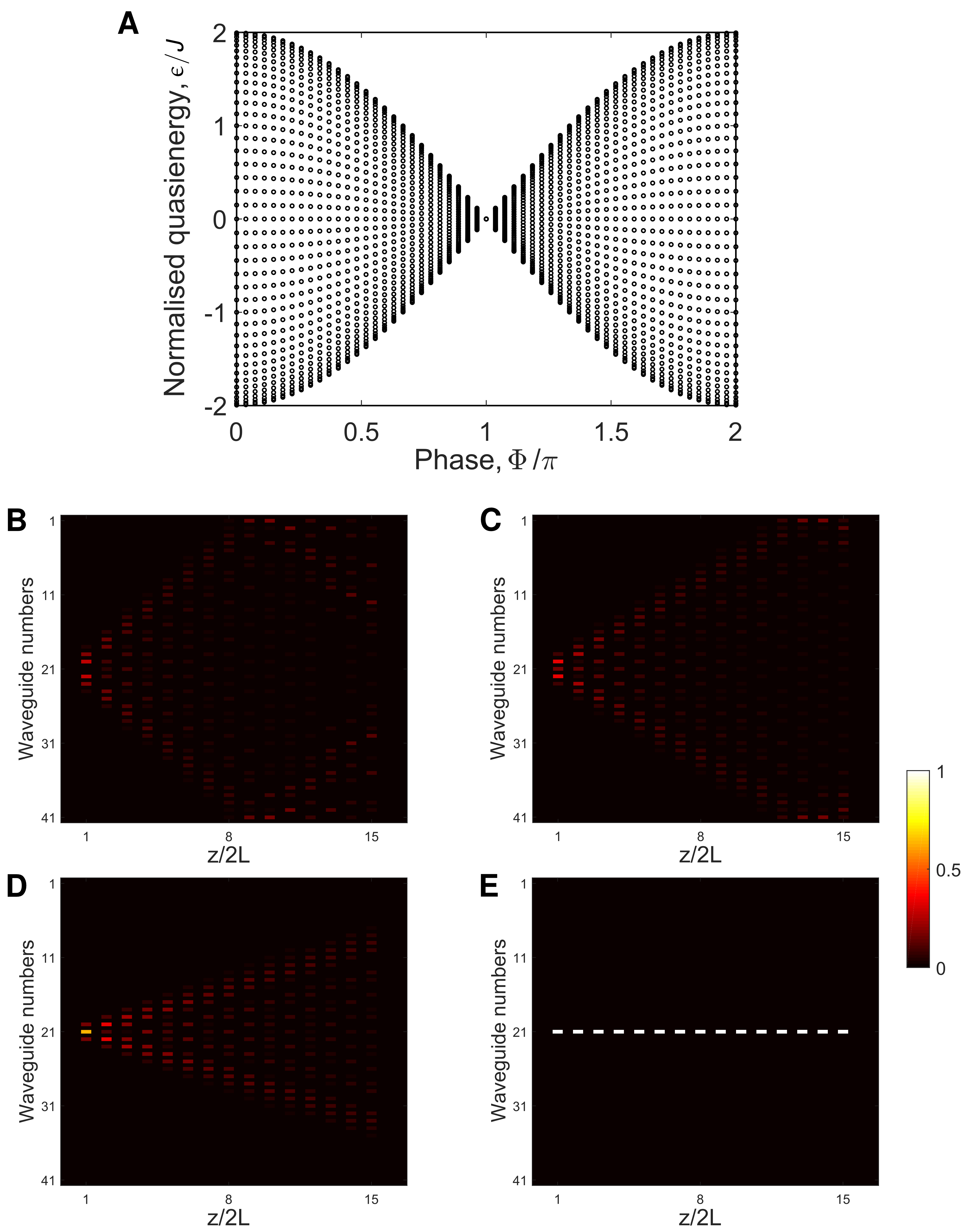}
\vspace{-0.cm} \caption{{\bf Floquet engineering of alternating pulsed electric field.} 
{\bf A,} Floquet quasienergy spectrum as a function of the phase shift between adjacent sites ($2\phi=\Phi$). The band collapses at $\Phi\!=\!\pi$ exhibiting destruction of tunneling. {\bf B-E,} Numerically calculated evolution of the intensity distributions for $\Phi\!=\!0, \pi/2, 3\pi/4$ and $\pi$ respectively. 
\label{DL}}
\end{figure}

\clearpage
\section*{\centering{ Supplementary Note 1: A one-dimensional driven lattice and the Dirac Hamiltonian}}

In this section, we present the driving protocol related to the 1D lattice illustrated in Figure 1B (in the main text) as well as its photonic implementation (Figure 1C). Let us first consider a general situation, a 1D tight-binding lattice with staggered hopping amplitudes ($J_{1,2}$). In the static case, the $k$-space Hamiltonian can be written as 
\begin{eqnarray}
\hat{H}_k\!=\! (J_1+J_2) \cos(kd) \hat{\sigma}_x-(J_1-J_2) \sin(kd) \hat{\sigma}_y,
\end{eqnarray}
where $d$ is the inter-site separation, $\hat{\sigma}_{x,y,z}$ are Pauli matrices and the Brillouin zone spans $0\! \le \!k\! \le \!\pi/d$. Now consider that $J_{1,2}$ are varying in a time periodic manner with a period $T$.  The driving protocol is the following:
\begin{eqnarray}
J_1\!=\!0 \ \text{and} \ J_2\!=\!\pi/T \quad \text{for} \ 0\le t \le T/2; \label{driving-1}\\
J_1\!=\!\pi/T \ \text{and} \ J_2\!=\!0 \quad \text{for} \ T/2\le t \le T. \label{driving-2}
\end{eqnarray}
Note that the probability for a particle to hop to its nearest site is unity when the tunneling/hopping is allowed. The Floquet operator and the effective Hamiltonian for this two-step driving protocol can be written as~\cite{Goldman2014periodically}
\begin{gather}
\hat{U}(T)\!=\! e^{-i\hat{H}_2T/2} e^{-i\hat{H}_1T/2} \nonumber\\
=\!\exp[-i(-2kd\hat{\sigma}_z+\pi \hat{1})]\!=\! e^{-i\hat{H}_{\text{eff}}T}, \label{Floquet-operator}\\
\hat{H}_{\text{eff}}= v_D k\hat{\sigma}_z+\text{cst}, \qquad v_D=-2d/T, \label{Heff}
\end{gather}
where $\hat H_{1,2}$ are the Hamiltonians for the two driving steps and $\hat{1}$ is a $2\times2$ identity matrix. 
Equation~\eqref{Heff} is the 1D Dirac Hamiltonian, which describes pseudo-relativistic particles in linearly-dispersive bands, with a ``speed of light" $v_D\!=\!2d/T$. The Floquet spectrum associated with Equation~\eqref{Heff} is shown in Figure 1D (main text).

Now let us discuss the photonic implementation of the driving protocol in Equation~\eqref{driving-1} and \eqref{driving-2}. To perform the state-recycling using a linear cavity, we consider a driven photonic lattice of length $L$ (as shown in Figure 1C) where each bond is a $50\!:\!50$ directional coupler. Note that light travels along the $+z$ direction for the first half of the complete driving period ($T\!\equiv\!2L$) and then almost $90\%$ of light reflects back and travels along the $-z$ direction. For the time-correlated single photon counting (TCSPC) measurement, the transmitted light is imaged onto the SPAD array. The four-step driving protocol for the photonic lattice can be written as follows: \\
{\text {Light propagation along $+z$ where}} 
\begin{eqnarray}
J_1\!=\!0 \ \text{and} \ J_2\!=\!\pi/(2L) \quad \text{for} \ 0\le z \le L/2; \label{driving-a}\\
J_1\!=\!\pi/(2L) \ \text{and} \ J_2\!=\!0 \quad \text{for} \ L/2\le z \le L; \label{driving-b}
\end{eqnarray}
{\text {and after reflection (i.e~light propagation along $-z$)}} 
\begin{eqnarray}
J_1\!=\!\pi/(2L) \ \text{and} \ J_2\!=\!0 \quad \text{for} \ L\le z \le L/2; \label{driving-c}\\
J_1\!=\!0 \ \text{and} \ J_2\!=\!\pi/(2L) \quad \text{for} \ L/2\le z \le 0. \label{driving-d}
\end{eqnarray}
Note that the propagation distance is the analogous time ($z\!\leftrightarrow\!t$). 
The Floquet operator for the four-step driving is given by 
\begin{eqnarray}
&\hat{U}(T)\!=\! \mathcal{T} \sum_n e^{-i\hat{H}_n T/4} \nonumber \\
&=\!\exp(-i[2kd(\sin(kd)\hat{\sigma}_x-\cos(kd)\hat{\sigma}_y)+\pi \hat{1}]), \label{FO-4steps}
\end{eqnarray}
where $\mathcal{T}$ indicates the time ordering and $n\!=\!1, 2, 3, 4$. The effective Hamiltonian then becomes
\begin{eqnarray}
\hat{H}_{\text{eff}}=(2d/T)k(\sin(kd)\hat{\sigma}_x-\cos(kd)\hat{\sigma}_y)+(\pi/T)\hat{1}. \label{Heff-4steps}
\end{eqnarray}
It can be shown that the effective Hamiltonian in Eq.~\eqref{Heff-4steps} is equivalent to the Dirac Hamiltonian in Eq.~\eqref{Heff}, up to a unitary transformation. The same result could be simply obtained by noting that the sequence \eqref{driving-a}-\eqref{driving-d} is equivalent to that of \eqref{driving-1}-\eqref{driving-2}, but in a different time frame.


\section*{\centering{ Supplementary Note 2: Brief description of anomalous topological edge modes}}

Topological band theory of a static system can be extended to a time periodic system using Floquet theory. In the presence of a high frequency driving, i.e.~driving frequency $\gg$ inter-site coupling strength, the topology of the system can be predicted by the usual topological invariants (e.g.~Chern numbers) that are used for a static system~\cite{Hasan2010}. However, away from the limit of high frequency driving, i.e.~driving frequency $\sim$ inter-site coupling strength, topologically protected edge modes can be observed even if the Chern numbers associated with all the bulk bands are zero. Such anomalous topological edge modes can be characterized by a distinct topological invariant known as the winding number~\cite{Rudner2013}.

Here we consider a square lattice with nearest neighbor coupling $J_{1-4}$ which are varying spatially and in a time periodic manner as shown in supplementary Figure~\ref{fig_driving_protocol}A. Imagine that the driving period, $T$, is equally split into four steps and $J_i T/4\!=\!\Lambda_i$. The Floquet operator for this driven lattice is given by 
\begin{equation}
U(T)\!=\!e^{-iH_4T/4}e^{-iH_3T/4}e^{-iH_2T/4}e^{-iH_1T/4},
\end{equation}
where the Hamiltonian, $\hat H_n$ ($n\!=\!1,2,3,4$), is piece-wise constant in time within the interval $(n-1)T/4\le t \le nT/4$, see supplementary Figure~\ref{fig_driving_protocol}A. Now consider that the first bond [step 1 in Figure~\ref{fig_driving_protocol}A] is always off i.e.~$\Lambda_1\!=\!0$ and for other three bond $\Lambda_{2-3}\!=\!\pi/2$.  For this driving protocol, the Floquet spectrum consists of two non-degenerate flat bands with zero Chern number, see supplementary Figure~\ref{fig_driving_protocol}B. Importantly, as the bulk bands are well separated, small disorder (less than the energy gap between the bulk bands) cannot close the energy gap. The winding numbers associated with both the energy gaps (centered on $0$ and $\pi/T$) are one. It should be highlighted that the magnitude of the group velocity of the chiral edge modes  along the top-right edge of the lattice (i.e.~along $y$ direction) is twice of that along the top-left (i.e.~along $x$ direction).

\section*{\centering{ Supplementary Note 3: Floquet engineering of an additional electric field}}

In this section, we show how the state-recycling technique can allow for the engineering of additional fields, which can then act on top of the (effective) Hamiltonian associated with the photonic lattice. Here, we illustrate this concept by showing how an effective electric field can be simply generated by modifying the very end of the lattice (which corresponds to pulsing a synthetic electric field after each round trip). Before describing this method, let us point out that such pulsed fields could also be digitally engineered, by changing the output (recycled) state according to a well-defined unitary operator. 

Let us consider a straight 1D photonic lattice with inter-waveguide separation, $d$ and nearest-neighbor coupling, $J$.  In the gray section illustrated in supplementary Figure~\ref{angled-lattice}A, the optical fields travel and exhibit discrete diffraction for a length $L$. The output facet of the lattice is polished at a small angle ($\theta$), hence the inter-waveguide light transfer in the red section  can be ignored for an experimentally realizable finite array. In this red section, the optical mode at the $s$-th waveguide will acquire an additional phase, $\phi_s\!=\!s(2\pi d n_{\text{eff}}/\lambda)\tan(\theta)\!=\!s\phi$ where $\lambda$ is the wavelength of light, $n_{\text{eff}}$ is the effective modal refractive index and $\phi$ is the inter-waveguide phase shift. It should be noted that this phase shift along the array, which is linear in $s$, is analogous to an external electric field applied instantaneously for an infinitesimally short duration of time; formally, this corresponds to acting on a state with the unitary operator $e^{-i \tau E \hat x}$, where $\tau$ is the effective pulse duration, $E=\phi/\tau$ is the effective electric field strength, and where $\hat x$ denotes the position operator on the lattice. Supplementary Figure~\ref{angled-lattice}B shows the variation of $\phi$ as a function of wavelength, providing an extra degree of freedom to fine-tune the strength and direction of the electric field.
When this lattice is placed inside a {\it ring cavity}, the complete driving period, $T=L$, consists of the following two steps:\\
(1) time evolution determined by the coupling strength, $J$ for $0\! \le \! z \! \le \! L$; \\
(2) tunneling is frozen and an analogous static field is applied at $z\!=\!L$.\\
Formally, the time-evolution operator over each period $T$ can be written in the two-step form,
\begin{equation}
\hat U (T)= e^{i (\phi/d) \hat x} \times e^{- i T \hat H_0 } , 
\end{equation}
where $\hat x\!=\!d\sum_s s \vert s \rangle \langle s \vert$ is the ``position" operator on the lattice and $\hat H_0=J \sum_s \vert s+1 \rangle\langle s \vert + \text{h.c.}$ is the hopping Hamiltonian with amplitude $J$. 
For weak synthetic electric fields, $\phi\!\ll\!1$, this time-evolution operator can be simplified according to the Trotter formula, 
\begin{equation}
\hat U (T)\approx e^{-i T \left (\hat H_0 - \phi/(dT) \, \hat x \right) } , \label{electric_effective}
\end{equation}
which describes the motion of a particle hopping on a lattice in the presence of an effective electric field $E=\phi/dT$. We note that a similar form can be obtained in the large-field regime $\phi\sim 1$, using the full Baker-Campbell-Hausdorff formula.

For the purpose of numerical calculations, a photonic lattice with $41$ waveguides and $\phi\!=\!\pi/7.5$ is considered. By solving the Schr\"odinger equation associated with the full Floquet (effective) Hamiltonian, the time evolution of the input state is obtained for two specific input states, see supplementary Figure~\ref{BO}. 
Supplementary Figure~\ref{BO}A corresponds to the input state localized at the $21$-st waveguide whereas for supplementary Figure~\ref{BO}B, seventeen waveguides were initially excited with a Gaussian intensity pattern. In both cases, the characteristics of Bloch oscillations are observed-- breathing motion of the intensity pattern for the first case and the oscillation of the wave-packet's center of mass for the second one.  

Now let us consider a different scheme generating alternating electric field pulses, which can be realized using a straight photonic lattice inside a {\it linear cavity} with both facets polished at equal angles, $\theta$. 
In this case, the complete driving period, $T\!=\!2L$, can be split into the following steps:\\
(1) time evolution determined by the coupling strength, $J$ for $0\! \le \! z \! \le \! L$; \\
(2) tunneling is frozen and an analogous static field (characterized by $2\phi$) is applied at $z\!=\!L$; \\
(3) time evolution determined by the coupling strength, $J$ for $L\! \le \! z \! \le \! 0$; \\
(4) tunneling is frozen and an analogous static field (characterized by $-2\phi$) is applied at $z\!=\!0$. \\
Formally, the time-evolution operator over each period $T$ can be written in the four-step form,
\begin{equation}
\hat U (T)= e^{-i (2\phi/d) \hat x} \times e^{- i (T/2) \hat H_0 }\times e^{i (2\phi/d) \hat x} \times e^{- i (T/2) \hat H_0 },\label{Floquet_op_bis}
\end{equation}
where we used the same notations as above. Using the following expression~\cite{Goldman2014periodically},
\begin{equation}
e^{i \alpha \hat x} \hat H_0 e^{-i \alpha \hat x}=\hat H_0 \cos (\alpha) - \hat H_1 \sin (\alpha) ,  
\end{equation}
where $\hat H_1=-i J \sum_s ( \vert s+1 \rangle\langle s \vert - \text{h.c.})$, and noting that $[\hat H_1,\hat H_0]\!=\!0$, we finally obtain an exact form for the Floquet operator in Eq.~\eqref{Floquet_op_bis} with
\begin{equation}
\hat U (T)=e^{-i T \cos (\phi) \hat H_0},\label{Floquet_op_tri}
\end{equation}
which indicates that the pulsed and alternating electric field simply renormalizes the hopping amplitude $J\!\rightarrow\!J \cos(\phi)$. We note that the effective Hamiltonian appearing in Eq.~\eqref{Floquet_op_tri}, namely, $\hat H_{\text{eff}}\!=\!\cos (\phi) \hat H_0$, is valid for any $\phi$. In particular, for $\phi\!=\!\pi/2$, we find that  the Floquet operator is trivial, $\hat U (T)\!=\!\hat{1}$, which indicates that the hopping is effectively annihilated by the pulsed electric field.

It should be mentioned that this particular four-step driving protocol can also be realized by fabricating the photonic lattice such that the waveguide axes are tilted at an angle with respect to the length of the glass sample. In that case, both facets will be at equal angles with respect to the waveguide axes without the requirement of angle polishing. 
For this driving protocol, the Floquet quasienergy spectrum as a function of the phase shift between adjacent sites ($2\phi\!=\!\Phi$) is presented in supplementary Figure~\ref{DL}A. As expected, the band indeed collapses at $\Phi\!=\!\pi$, indicating destruction of tunneling. Supplementary Figure~\ref{DL}B-E show numerically calculated evolution of the intensity distributions for $\Phi\!=\!0, \pi/2, 3\pi/4$ and $\pi$ respectively.

\section*{\centering{ Supplementary Note 4: Discrete diffraction in the presence of a synthetic electric field}}

In this section, we present experimental details and the driving protocol related to the quasi-real time-resolved imaging of discrete diffraction in the presence of a synthetic electric field; see Figure~2 in the main text. Using ultrafast laser inscription, a 1D straight photonic lattice consisting of twenty coupled single-mode waveguides was fabricated. Both facets of the glass substrate containing the lattice were polished and silver-coated to form a linear cavity. The input and output facet angles with respect to the lattice axis ($x$) were measured to be $\theta_{1, 2}\!\approx\! \pm 0.1^{\circ}$, 
respectively. These small angles at the facets of the substrate (inset in Fig.~2E, main text)
cause a linear phase shift along the lattice, which effectively produces a time-periodic (pulsed) synthetic electric field.
In contrast to our previous discussion in Eq.~\eqref{Floquet_op_bis}, here the angles have opposite signs, which means that the associated synthetic pulsed electric fields will be along the same direction.   
In this experiment, the total driving period, $T\!=\!2L$, consists of four driving steps:\\
(1) tunneling is frozen and an analogous static field (characterized by $+2\phi_1$) is applied at $z\!=\!0$; \\
(2) time evolution determined by the coupling strength, $J$ for $0\! \le \! z \! \le \! L$; \\
(3) tunneling is frozen and an analogous static field (characterized by $+2\phi_2$) is applied at $z\!=\!L$; \\
(4) time evolution determined by the coupling strength, $J$ for $L\! \le \! z \! \le \! 0$; \\
Formally, the time-evolution operator over each period $T$ can be written in the four-step form,
\begin{eqnarray}
\hat U (T)&= e^{- i (T/2) \hat H_0 }
\times e^{i (2\phi_2/d) \hat x} 
\times e^{- i (T/2) \hat H_0 }
\times e^{i (2\phi_1/d) \hat x} \label{Floquet_BO_expt} \\
&\approx e^{-i \, T \big(\hat H_0 - 2(\phi_1+\phi_2)/(d\,T) \, \hat x \big) }, \label{Floquet_BO_expt2}
\end{eqnarray}
where $\hat H_0$ describes the hopping in the 1D array, see Eq.~\eqref{electric_effective}. Eq.~\eqref{Floquet_BO_expt2} describes the motion of a particle hopping on a lattice in the presence of an effective electric field $E\!=\!2(\phi_1+\phi_2)/(dT)$. 
In our experiment, the expected inter-waveguide phase shifts are $\phi_{1,2}\!\approx\!\pi/12$ which means that the associated Bloch period is $\approx\!12L\!=\!360 \ $mm. In Figure 2 (main text), we probed the dynamics up to $210 \ $mm which is approximately half of this period.

In conclusion, the state-recycling technique introduced in this work indeed allows one to engineer synthetic electric fields, which effectively act on top of the dynamics associated with the engineered photonic lattice. For constant electric field [Eq.~\eqref{electric_effective}], this could be used to perform transport experiments in view of probing response functions (e.g.~the conductivity tensor) or geometric effects through Bloch oscillations~\cite{Wimmer2017}. For alternating electric fields, this could be used to control the properties (e.g.~tunneling) of the engineered photonic lattice.
\vspace*{0.5cm}

\end{document}